\documentclass[
amssymb,nobibnotes,aps,pre]{revtex4}

\usepackage{hyperref}
\usepackage{graphicx}
\usepackage{color}
\usepackage{bm}
\usepackage{graphicx,amssymb,amstext,amsmath}
\usepackage{epsfig}
\usepackage{epstopdf}
\usepackage{color,graphics}

\begin{document}

\title{Systematic Derivation of Noether Point Symmetries \\ in Special Relativistic Field Theories}

\author{Fernando Haas}

\affiliation{Physics Institute, Federal University of Rio Grande do Sul, Avenida Bento Gon\c{c}alves 9500, 91501-970 Porto Alegre, RS, Brazil}

\date{\relax}

\begin{abstract}
A didactic and systematic derivation of Noether point symmetries and conserved currents is put forward in special relativistic field theories, 
without {\it a priori} assumptions about the transformation laws. Given the Lagrangian density, the invariance condition develops as a set of partial differential equations determining the symmetry transformation. The solution is provided in the case of real scalar, complex scalar, free electromagnetic, and charged electromagnetic fields. Besides the usual conservation laws, a less popular symmetry is analyzed: the symmetry associated with the linear superposition of solutions, whenever applicable. The role of gauge invariance is emphasized. The case of the charged scalar particle under external electromagnetic fields is considered, and the accompanying Noether point symmetries determined.
\end{abstract}

\maketitle



\section{Introduction}

Symmetry is a fundamental concept in Physics. In the Lagrangian formalism, Noether's symmetries play a central role, since Noether's theorem shows a direct connection between symmetries and conservation laws \cite{Noether}. In particular, the starting point of modern gauge theories \cite{Kleinert, Ryder} is a Lagrangian density admitting a prescribed symmetry group. The associated conservation laws should then reflect experimental observations. Examples of conserved quantities include energy, linear momentum, angular momentum, and electric charge. 

A natural requirement in relativistic theories is that the Lagrangian density must be a Lorentz scalar. This assures the validity of the Poincar\'e group, which is the fundamental transformation group on Minkowski space. In another context, in gauge theories, interactions are obtained from local gauge invariance principles. Usually, the local gauge symmetry is postulated in advance. Afterward, the action invariance is explicitly verified \cite{Kleinert, Ryder}. The question, regarding an opposite point of view, is how to systematically derive Noether symmetries uniquely from the Lagrangian density, without additional hypothesis. The advantage of the deductive approach is the possibility of unveiling symmetries not apparent from the very beginning. In addition, given a broad class of Lagrangians, one might be interested in the identification of the particular subclasses admitting Noether symmetries. In Section 7, this approach is put forward in the case of the charged scalar particle under prescribed external electromagnetic fields. In this case, not necessarily the external field will be compatible with some symmetry transformation. It should be mentioned that the traditional approach, put forward e.g. in authoritative textbooks 
\cite{Kleinert, Ryder, Schweber, Weinberg} is of course perfectly well justified, as long as the overall symmetry structure of space-time and internal space is concerned. However, sometimes (as in the aforementioned case of the charged scalar particle under external fields) one might be interested in subclasses admitting Noether point symmetry, as a toll in the search for sufficiently simple benchmark systems. 

The present article shows in detail the procedure for the systematic derivation of Noether point (or geometric) symmetries, applied to special relativistic field theories. In fact, in comparison with systems of a discrete number of degrees of freedom, there are fewer examples of 
step-by-step calculation of Noether invariance results for continuous systems, in relativity. For instance, Ref. \cite{Kobussen} considers the derivation of integrals of motion for the N-body problem. With few exceptions \cite{Havas, Schoeller}, most works in this direction assume non-relativistic and discrete systems.

The work considers some of the basic relativistic field theories, going from the simplest to the more elaborate models. Namely, in a sequence, the real scalar, complex scalar, vacuum electromagnetic and coupled complex scalar and electromagnetic field theories will be treated, as case examples for the systematic application of Noether's theorem in the search for geometric transformations. In this way, an hierarchy of models will be described from the symmetry analysis point of view. Although in all studied cases the external symmetries are given by the Poincar\'e group, as expected, the internal symmetries have distinctive features, including global or local gauge symmetries, together with an additional internal symmetry due to linearity, in the non-interacting field cases - to be detailed in the next sections. It is assumed that the present communication has essentially a methodological character. Nevertheless, it is precisely the systematic (non {\it ad hoc}) procedure that allows the identification of the aforementioned extra internal symmetry. Moreover, it will be addressed the case of a charged scalar particle under external electromagnetic fields. The class of external fields so that Noether point symmetries are admissible will be so determined, for the first time, as well as the accompanying conservation laws. The problem has interest for strong laser-plasma interactions, where test-particle dynamics has both relativistic and quantum aspects \cite{Marklund}.

Evidently, nowadays a large number of packages is available for the calculation of Noether point symmetries using computer algebra software. Nevertheless, the interest still remains for some people at least, to not blindly follow the computer's advice and to personally understand all the steps in the symmetry procedure. Moreover, in the same trend, the version of the Noether theorem presented below, is certainly not the most abstract, general or rigorous possible. Nevertheless, for practical applications, it might be of some interest to present the subject in a modest, more readable fashion for non-mathematicians. Finally, it should be noted that the manuscript is not intended to be a review. Therefore a detailed, encyclopaedic account of updated references on symmetries and conservation laws should be found elsewhere. 

This work is organized as follows. In Section 2, the Noether theorem is reviewed. In Section 3, the Noether point symmetries and conservation laws are deduced in the case of the real scalar field. The same procedure is repeated in the remaining sections, for increasing complexity of the models. Section 4 is dedicated to the complex scalar field. Section 5 is dedicated to the vacuum electromagnetic field. Section 6 considers the coupled complex scalar and electromagnetic fields. Section 7 is devoted to the charged scalar particle under external electromagnetic fields. Section 8 presents some conclusions. 

\section{Noether's Theorem}
\label{Noethertheorem}

In this Section, the Noether's theorem is enunciated in the case of a single field, the generalization to the multi-field case being straightforward. The starting point is the action functional, 
\begin{equation}
\nonumber
S = \int\,{\cal{L}}(\phi,\partial_{\mu}\phi,x)\,d^{4}x \,,
\end{equation}
where $\phi = \phi(x)$ is the pertinent field, ${\cal L}$ is the Lagrangian density of the indicated arguments, and $x$ is a 4-vector with covariant components $x_\mu = (t, {\bf r}) = (x_0, x_1, x_2, x_3)$. The metric tensor will be taken as
$g^{\mu\nu} = {\rm diag}(1,-1,-1,-1)$. Natural units $c =  1, \hbar = 1$ and the Einstein summation convention will be employed. Greek indexes run from 0 to 3, and Latin indexes from 1 to 3.

The infinitesimal point transformations given by 
\begin{eqnarray}
x^\mu &\rightarrow& x^{\mu} + \varepsilon\,\eta^{\mu}(\phi,x) \,, \nonumber \\
\phi &\rightarrow& \phi + \varepsilon\,\psi(\phi,x) \,, \nonumber
\end{eqnarray}
will be considered, where $\varepsilon$ is a real infinitesimal parameter and where
$\eta^{\mu}(\phi,x)$, $\psi(\phi,x)$ are smooth functions not dependent on the field derivatives. Dynamical symmetries, where the transformation law involves the field derivatives, are fundamental in many cases \cite{Bluman}. 
For instance, the derivation of the infinite hierarchy of conserved functionals for the Korteweg-de Vries equation \cite{Olver} needs the application of dynamical symmetries. Nevertheless, for simplicity this work is restricted to point transformations only. 

The (quasi) invariance condition for the action is given \cite{Hill, Sarlet} by 
\begin{equation}
\label{eq1} \frac{\partial {\cal{L}}}{\partial\phi}\,\psi +
\frac{\partial
{\cal{L}}}{\partial(\partial_{\mu}\phi)}\,(d_{\mu}\psi
-
d_{\nu}\phi\,\partial_{\mu}\eta^{\nu}) +
\partial_\mu{\cal{L}}\,\,\eta^\mu +
{\cal{L}}\,\,d_{\mu}\eta^\mu = d^{\mu}\sigma_\mu \,,
\end{equation}
where $\sigma_\mu = \sigma_{\mu}(\phi,x)$ is at this stage an arbitrary 4-vector of the indicated arguments. Actually, the condition 
(\ref{eq1}) does not assures the strict invariance of the action, which can be modified by the addition of a constant surface term, namely, the surface integral of $\sigma_\mu$ at infinity. However, a numerical constant added to the action, has not any effect on the form of the Euler-Lagrange equations.  

For the sake of notation, we denote total derivatives as $d_\mu$ and partial derivatives (maintaining constant fields and derivatives of the fields) as 
$\partial/\partial\,x^\mu = \partial_\mu$. For instance, 
\begin{equation}
d_{\mu}{\cal{L}} = \partial_\mu{\cal{L}} +
\frac{\partial {\cal{L}}}{\partial\phi}\,\partial_{\mu}\phi +
\frac{\partial
{\cal{L}}}{\partial(\partial_{\nu}\phi)}\,\partial_{\mu}\partial_{\nu}\phi
\,. \nonumber
\end{equation}

Noether's theorem assures that whenever the symmetry condition (\ref{eq1}) is satisfied, there is a conserved current given by 
\begin{equation}
\label{eqq1} J^\mu = \theta^{\mu}_{\,\nu}\,\eta^\nu - \frac{\partial
{\cal{L}}}{\partial(\partial_{\mu}\phi)}\,\psi + \sigma^\mu \,,
\end{equation}
where the energy-momentum tensor $\theta^{\mu}_\nu$ is defined by 
\begin{equation}
\nonumber
\theta^{\mu}_{\,\nu} = \frac{\partial
{\cal{L}}}{\partial(\partial_{\mu}\phi)}\,\partial_{\nu}\phi -
\delta^{\mu}_{\,\nu}\,{\cal{L}} \,.
\end{equation}
The conservation law reads 
\begin{equation}
\nonumber
d_{\mu}J^\mu = 0 \,,
\end{equation}
where $\phi$ solves the Euler-Lagrange equation, 
\begin{equation}
\label{el}
\frac{\partial{\cal L}}{\partial\phi} - d_{\mu}\left(\frac{\partial{\cal L}}{\partial(\partial_{\mu}\phi)}\right) = 0 \,.
\end{equation}

The basic question to be addressed here is: given the Lagrangian density, how to systematically derive the infinitesimal symmetry transformations leaving the action functional invariant up to the addition of a surface term? The answer, to be developed in the examples in the next sections, is as follows. Inserting the Lagrangian density into the invariance condition (\ref{eq1}), typically we obtain a polynomial equation on the field derivatives. The coefficient of each different field derivative must be zero. Otherwise, one would impose additional constraints on the field, which should be ideally leaved free as much as possible. Therefore, a set of partial differential equations for the symmetry functions will be derived, not involving the derivatives of the field. Solving the determining partial differential equations in all generality, we obtain the full set of Noether point symmetries, without {\it ad hoc} postulates. The procedure will be worked out in the following examples, starting with one of the simplest relativistic models, namely, the Klein-Gordon field. 

\section{Real Scalar Field}
\label{realscalar}

\subsection{Noether Symmetries for the Real Scalar Field}
The Lagrangian density for the real scalar field is
\begin{equation}
\label{lsf}
{\cal{L}} = \frac{1}{2}\,\partial^{\mu}\phi\,
\partial_{\mu}\phi - \frac{1}{2}\,
m^2\,\phi^2 \,,
\end{equation}
where $m$ is the particle mass. Using the Euler-Lagrange equation (\ref{el}), we derive Klein-Gordon's equation, 
\begin{equation}
\nonumber
\partial^{\mu}\partial_{\mu}\phi + m^{2}\phi = 0 \,.
\end{equation}

Inserting ${\cal{L}}$ from Eq. (\ref{lsf}) into the symmetry condition (\ref{eq1}),
it follows that
\begin{equation}
\label{equa2} - m^{2}\phi\,\psi +
\partial^{\mu}\phi\,(d_{\mu}\psi -
\partial_{\nu}\phi\,d_{\mu}\eta^{\nu}) +
\frac{1}{2}\,(\partial^{\mu}\phi\,\partial_{\mu}\phi -
m^{2}\phi^2)d_{\nu}\eta^\nu = d^{\mu}\sigma_\mu \,.
\end{equation}
The quantities $\eta^\mu$ and $\psi$ should be managed so that the left-hand side of Eq. (\ref{equa2}) becomes the divergence of some appropriate 4-vector $\sigma_\mu$.

The following total derivatives
\begin{eqnarray} \nonumber
d_{\mu}\psi &=& \partial_\mu\psi +
\frac{\partial\psi}{\partial\phi}\,\partial_{\mu}\phi \,,\\
\nonumber
d_{\mu}\eta^\nu &=& \partial_\mu\eta^\nu +
\frac{\partial\eta^\nu}{\partial\phi}\,\partial_{\mu}\phi \,,\\
\nonumber
d^{\mu}\sigma_\mu &=& \partial^\mu\sigma_\mu +
\frac{\partial\sigma_\mu}{\partial\phi}\,\partial^{\mu}\phi \,,
\end{eqnarray}
when inserted on Eq. (\ref{equa2}), give the expression 
\begin{eqnarray}
-
\frac{1}{2}\,\partial^{\mu}\phi\,\partial_{\mu}\phi\,\partial_{\nu}
\phi\,\frac{\partial\eta^\nu}{\partial\phi}
+ \partial^{\mu}\phi\,\partial_{\mu}\phi\,\left(\frac{\partial\psi}{\partial\phi}
+ \frac{1}{2}\,\partial_\nu\eta^\nu\right) -
\partial^{\mu}\phi\,\partial_{\nu}\phi\,\partial_\mu\eta^\nu + \nonumber
\\
\label{equa3}
\partial^{\mu}\phi\,\left(\partial_\mu\psi -
\frac{1}{2}\,m^{2}\phi^{2}\,\frac{\partial\eta_\mu}{\partial\phi}\right) -
m^{2}\phi\,\psi -
\frac{1}{2}\,m^{2}\phi^{2}\,\partial_\mu\eta^\mu =
\partial^{\mu}\phi\,\frac{\partial\sigma_\mu}{\partial\phi} +
\partial^\mu\sigma_\mu \,.
\end{eqnarray}

Equation (\ref{equa3}), to be identically satisfied, is a polynomial expression on the field derivatives.
Therefore, the coefficient of each monomial (term with equal derivative power) should vanish. The third order terms give 
\begin{equation}
\nonumber
\partial^{\mu}\phi\,\partial_{\mu}\phi\,\partial_{\nu}
\phi\,\frac{\partial\eta^\nu}{\partial\phi}
= 0 \quad \Rightarrow \quad \eta^{\mu} = \eta^{\mu}(x) \,.
\end{equation}
In another words, the external transformations (affecting the space-time coordinates only) are 
not dependent on $\phi$. 

The second order terms in Eq. (\ref{equa3}) imply 
\begin{equation}
\partial^{\mu}\phi\,\partial_{\mu}\phi\,\left(\frac{\partial\psi}{\partial\phi}
+ \frac{1}{2}\,\partial_\nu\eta^\nu\right) -
\partial^{\mu}\phi\,\partial_{\nu}\phi\,\partial_\mu\eta^\nu = 0 \,.
\nonumber
\end{equation}
The last equation decomposes itself into a set of equations, corresponding to terms proportional to 
$\partial_{0}\phi\,\partial_{0}\phi$, $\partial_{0}\phi\,\partial_{i}\phi$ and
$\partial_{i}\phi\,\partial_{j}\phi$. A detailed examination shows that the resulting equations are reducible to 
\begin{eqnarray}
\label{equa4} - \partial_{0}\eta_0 &=& \partial_{1}\eta_1 =
\partial_{2}\eta_2 = \partial_{3}\eta_3 \,,\\ \label{equa5}
\partial_{\mu}\eta_\nu &+& \partial_{\nu}\eta_\mu = 0 \,,\quad \mu \neq \nu
\,,
\end{eqnarray}
together with
\begin{equation}
\label{equa6}
\frac{\partial\psi}{\partial\phi} = - \partial_{0}\eta_0 \,,
\end{equation}
where Eq. (\ref{equa4}) was taken into account for Eq. (\ref{equa6}). 
We left the system (\ref{equa4})-(\ref{equa5}) untouched by now. Equation (\ref{equa6}) gives 
\begin{equation}
\label{equa7} \psi = - \phi\,\partial_{0}\eta_0 + \tilde\phi(x)
\,,
\end{equation}
where $\tilde\phi(x)$ is an arbitrary function of $x$.

The invariance condition (\ref{equa3}), for the terms which are of first order in the derivatives, leaves us with 
\begin{equation}
\frac{\partial\sigma_\mu}{\partial\phi} = \partial_\mu\psi = -
\phi\,\partial_{\mu}\partial_{0}\eta_0 + \partial_{\mu}\tilde\phi \,, 
\nonumber
\end{equation} 
with the solution
\begin{equation} 
\label{eq8}
\sigma_\mu = - \frac{\phi^2}{2}\,\partial_{\mu}\partial_{0}\eta_0
+ \phi\,\partial_{\mu}\tilde\phi + \tilde\sigma_{\mu}(x) \,,
\end{equation}
where $\tilde\sigma_{\mu}(x)$ is an arbitrary 4-vector depending only on $x$.

The zeroth-order term on Eq. (\ref{equa2}) implies 
\begin{equation}
\label{zo}
- m^{2}\phi\,\psi - \frac{1}{2}\,m^{2}\phi^{2}\,\partial_\mu\eta^\mu =
\partial^\mu\sigma_\mu \,.
\end{equation}
Inserting the results from Eqs. (\ref{equa4}), (\ref{equa7}) and (\ref{eq8}) into Eq. (\ref{zo}), one get  
\begin{equation}
\label{ee}
\left(- \frac{1}{2}\,\partial^{\mu}\partial_{\mu}\partial_{0}\eta_0
+ m^{2}\partial_{0}\eta_0\right)\,\phi^2 +
\left(\partial^{\mu}\partial_{\mu}\tilde\phi +
m^{2}\tilde\phi\right)\,\phi + \partial^{\mu}\tilde\sigma_\mu = 0
\,.
\end{equation}
Equation (\ref{ee}), being identically satisfied for any $\phi$, implies that the coefficients of different powers of the field vanish, or, 
\begin{eqnarray}
\label{eq9} -
\frac{1}{2}\,\partial^{\mu}\partial_{\mu}\partial_{0}\eta_0 +
m^{2}\partial_{0}\eta_0 = 0 \,,\\ \label{eq10}
\partial^{\mu}\partial_{\mu}\tilde\phi + m^{2}\tilde\phi = 0 \,,\\
\label{eq11}
\partial^{\mu}\tilde\sigma_\mu = 0 \,.
\end{eqnarray}
Equation (\ref{eq9}) can be rewritten as 
\begin{equation}
\label{eq12}
\partial_{0}\left(\partial^{\mu}\partial_{\mu}\eta_0 - 2\,m^{2}\eta_0\right) = 0 \,,
\end{equation}
or, 
\begin{equation}
\label{eqq300} \partial^{\mu}\partial_{\mu}\eta_0 - 2\,m^{2}\eta_0 =
- 2\,m^{2}\tilde\eta_{0}({\bf r}) \,,
\end{equation}
where $\tilde\eta_{0}({\bf r})$ is a function of space coordinates only. 

The next information, comes from Eq. (\ref{eq10}) showing that 
$\tilde\phi$ solves the Klein-Gordon equation. Therefore, adding to $\phi$ a particular solution of the Klein-Gordon equation is a Noether symmetry, reflecting the linearity of the equation. In addition, Eq. (\ref{eq11}) shows that 
\begin{equation}
\nonumber
\tilde\sigma_\mu = 0 \,,
\end{equation}
without loss of generality. 

While Eqs. (\ref{eq10}) and (\ref{eq11}) have already been fully examined, there remains Eq.  (\ref{eq9}), which is equivalent to Eq. 
(\ref{eqq300}). To analyze the last one, we take into account Eqs. (\ref{equa4}) and (\ref{equa5}). For instance, considering the first line in Eq. (\ref{equa4}), for  $\mu = 0$, $\nu = 1$ in Eq. (\ref{equa5}), it results 
\begin{equation}
\nonumber
\partial_{1}\eta_1 = - \partial_{0}\eta_0 \,,\quad \partial_{0}\eta_1 = -
\partial_{1}\eta_0 \,.
\end{equation}
To satisfy Cauchy's condition $\partial_{0}\partial_{1}\eta_1 =
\partial_{1}\partial_{0}\eta_1$, necessarily 
\begin{equation}
\label{eq13}
\partial_{1}\partial_{1}\eta_0 = \partial_{0}\partial_{0}\eta_0 \,.
\end{equation}
Similarly, using again Eqs. (\ref{equa4}) and (\ref{equa5}),
we conclude that 
\begin{equation}
\label{eq14}
\partial_{2}\partial_{2}\eta_0 =
\partial_{3}\partial_{3}\eta_0 =
\partial_{0}\partial_{0}\eta_0 \,.
\end{equation}
Equations (\ref{eq13}) and (\ref{eq14}) allow to write 
\begin{equation}
\nonumber
\partial^{\mu}\partial_{\mu}\eta_0 = - 2\,\partial_{0}\partial_{0}\eta_0 \,.
\end{equation}
Inserting the last into Eq. (\ref{eq12}) gives
\begin{equation}
\label{eq15}
\partial_{0}\partial_{0}\eta_0 + m^{2}\eta_0 = m^{2}\tilde\eta_{0}({\bf r}) \,.
\end{equation}
Only time-derivatives appear in the differential equation (\ref{eq15}). Therefore, in this context,
$\tilde\eta_{0}({\bf r})$ is a constant, and the general solution obviously is 
\begin{equation}
\label{eq16} \eta_0 = \tilde\eta_{0}({\bf r}) + F({\bf r})e^{imt} +
G({\bf r})e^{- imt} \,,
\end{equation}
where $F({\bf r})$ and $G({\bf r})$ are arbitrary functions of the space coordinates only. 

From now on, sometimes we denote ${\bf r} = (x,y,z)$ whenever convenient, as long as there is no risk of confusion between the space-time 4-vector $x = (t, {\bf r})$ and the coordinate $x$. Following this definition, and inserting Eq. (\ref{eq16}) into (\ref{equa4}), the result is %
\begin{eqnarray}
\label{eq301} \eta_1 &=& \tilde\eta_{1}(y,z,t) -
im\,e^{imt}\int\,dx\,F + im\,e^{-imt}\int\,dx\,G \,,\\
\label{eq302} \eta_2 &=& \tilde\eta_{2}(x,z,t) -
im\,e^{imt}\int\,dy\,F + im\,e^{-imt}\int\,dy\,G \,,\\
\label{eq303} \eta_3 &=& \tilde\eta_{3}(x,y,t) -
im\,e^{imt}\int\,dz\,F + im\,e^{-imt}\int\,dz\,G \,,
\end{eqnarray}
where $\tilde\eta_1, \tilde\eta_2$ and $\tilde\eta_3$ are arbitrary functions of the indicated arguments. 

The solution presented in Eqs. (\ref{eq16})-(\ref{eq303}) must be compatible with Eq. (\ref{equa5}).
For instance, for $\mu = 0, \nu = 1$ in Eq. (\ref{equa5}), we obtain 
\begin{equation}
\label{eqq18}
\frac{\partial}{\partial\,t}\tilde\eta_{1} +
\frac{\partial}{\partial\,x}\tilde\eta_{0} +
e^{imt}\,\left(m^{2}\int\,dx\,F +
\frac{\partial\,F}{\partial\,x}\right) + e^{- imt}\,\left(m^{2}\int\,dx\,G +
\frac{\partial
G}{\partial\,x}\right) = 0 \,. \nonumber
\end{equation}
The derivative of the last equation with respect to $x$ implies 
\begin{equation}
\label{eq19}
\frac{\partial^{2}\tilde\eta_{0}}{\partial\,x^2} +
e^{imt}\left(\frac{\partial^{2}F}{\partial\,x^2} + m^{2}F\right) +
e^{- imt}\left(\frac{\partial^{2}G}{\partial\,x^2} + m^{2}G\right) \nonumber
= 0 \,.
\end{equation}
Since neither $\tilde\eta_{0}$, nor $F, G$ have a dependence on time, it follows that 
\begin{eqnarray}
\label{eq400}
\frac{\partial^{2}F}{\partial\,x^2} + m^{2}F &=&
\frac{\partial^{2}G}{\partial\,x^2} +
m^{2}G = 0 \,,\\ \label{eq401}
\frac{\partial^{2}\tilde\eta_{0}}{\partial\,x^2} &=& 0 \,.
\end{eqnarray}
Equation (\ref{eq401}) will not have immediate consequences. On the other hand, following a procedure similar to the derivation of Eq. 
(\ref{eq400}), we obtain 
\begin{eqnarray}
\label{eq402}
\frac{\partial^{2}F}{\partial\,y^2} + m^{2}F &=& \frac{\partial^{2}F}{\partial\,z^2} +
m^{2}F = 0 \,,\\ \label{eq403}
\frac{\partial^{2}G}{\partial\,y^2} + m^{2}G &=& \frac{\partial^{2}G}{\partial\,z^2} +
m^{2}G = 0 \,.
\end{eqnarray}
The (unique) solution for the system composed by Eqs. (\ref{eq400}), 
(\ref{eq402})-(\ref{eq403}) is
\begin{eqnarray}
F &=& c_{1}\exp[im(x+y+z)] + c_{2}\exp[im(x+y-z)]  \nonumber \\
&+& c_{3}\exp[im(x-y+z)] + c_{4}\exp[im(-x+y+z)]  \nonumber \\
&+& c_{5}\exp[im(x-y-z)] + c_{6}\exp[im(-x+y+z)] \nonumber \\ &+&
c_{7}\exp[im(-x-y+z)] + c_{8}\exp[-im(x+y+z)] \,, \nonumber \\ G &=&
c_{9}\exp[im(x+y+z)] + c_{10}\exp[im(x+y-z)] \nonumber \\
&+& c_{11}\exp[im(x-y+z)] + c_{12}\exp[im(-x+y+z)] \nonumber \\
&+& c_{13}\exp[im(x-y-z)] + c_{14}\exp[im(-x+y+z)] \nonumber \\ &+&
c_{15}\exp[im(-x-y+z)] + c_{16}\exp[-im(x+y+z)] \,, \nonumber
\end{eqnarray}
where $c_1,\dots\,c_{16}$ are numerical constants. Hence,   
\begin{eqnarray}
\label{equa500} \eta_0 &=& \tilde\eta_{0}(x,y,z) +
c_{1}\exp[im(x+y+z+t)]  \nonumber \\ &+& \dots +
c_{16}\exp[-im(x+y+z+t)] \,,\\ \label{eq502} \eta_1 &=&
\tilde\eta_{1}(y,z,t) - c_{1}\exp[im(x+y+z+t)]  \nonumber
\\ &+& \dots - c_{16}\exp[-im(x+y+z+t)] \,,\\ \label{eq503} \eta_2 &=&
\tilde\eta_{2}(x,z,t) - c_{1}\exp[im(x+y+z+t)]  \nonumber
\\ &+& \dots - c_{16}\exp[-im(x+y+z+t)] \,,\\ \label{eq501} \eta_3 &=&
\tilde\eta_{3}(x,y,t) - c_{1}\exp[im(x+y+z+t)]  \nonumber
\\ &+& \dots - c_{16}\exp[-im(x+y+z+t)] \,,
\end{eqnarray}
where the terms depending on $c_2,
\dots, c_{15}$ were omitted, for brevity.

It remains the constraint (\ref{equa5}). Taking $\mu = 1, \nu = 2$ in Eq. 
(\ref{equa5}) and inserting Eqs. (\ref{eq502})-(\ref{eq503}), results in 
\begin{eqnarray}
\partial_{1}\tilde\eta_2 + \partial_{2}\tilde\eta_1 &-&
2\,imc_{1}\exp[im(x+y+z+t)]  \nonumber \\ &+& \dots +
2\,imc_{16}\exp[- im(x+y+z+t)] = 0 \nonumber \,.
\end{eqnarray}
Derivation of the last with respect to $x$ and 
$y$, shows that
\begin{equation}
2\,im^{3}\left\{c_{1}\exp[im(x+y+z+t)] + \dots + c_{16}\exp[-
im(x+y+z+t)]\right\} = 0 \,, \nonumber
\end{equation}
so that 
\begin{equation}
c_1 = \dots = c_{16} = 0 \,. \nonumber
\end{equation}
Since all numerical constants $c_1,\dots,c_{16}$
vanish, and using Eqs. (\ref{equa5}), (\ref{equa500})-(\ref{eq501}),
we get the compact expressions
\begin{equation}
\label{eq600}
\partial_{\mu}\eta_\nu + \partial_{\nu}\eta_\mu = 0 \,,
\end{equation}
generalizing Eq. (\ref{equa5}) to arbitrary indexes $\mu, \nu$. 

We are almost done. However, we should still take into account Eq. (\ref{eq600}), applying appropriate derivatives to it. For instance, for 
$\mu = 0, \nu = 1$, differentiation with respect to $x$ recalling that $\tilde\eta_0$ does not depend on time, gives 
\begin{equation}
\partial_{1}\partial_{1}\eta_{0} = 0 \,. \nonumber
\end{equation}
Similar calculations, involving appropriate indexes and derivatives of Eq. (\ref{eq600}), shows that the functions $\eta^\mu$ are at most linear functions of space-time coordinates. In other words, 
\begin{equation}
\tilde\eta^\mu = a^\mu + R^{\mu}_{\,\nu}\,x^\nu \,, \nonumber
\end{equation}
where $a^\mu$ is an arbitrary constant 4-vector and $R^{\mu}_{\,\nu}$ is a constant second-rank tensor.  However, the components 
$R^{\mu}_{\,\nu}$ are not entirely free. Indeed, Eq. (\ref{eq600}) implies 
\begin{equation}
\nonumber
R_{\mu\nu} + R_{\nu\mu} = 0 \,,
\end{equation}
so that $R^{\mu}_{\,\nu}$ is an anti-symmetric tensor. This exhausts the information contained in the 
Noether symmetry condition (\ref{eq1}), applied to the real scalar field. 

We are at a convenient point to enumerate the results until now. The Noether symmetries are completely specified by 
\begin{eqnarray}
\label{poi}
\eta^\mu &=& a^\mu + R^{\mu}_{\,\nu}\,x^\nu \,, \nonumber \\ \psi &=&
\tilde\phi(x) \,, \nonumber
\end{eqnarray}
where $a^\mu$ is a constant 4-vector and $R^{\mu}_{\,\nu}$ a constant anti-symmetric tensor, while $\tilde\phi$ is any solution of the Klein-Gordon equation. Therefore, we are left with a 10-parameter external symmetry group for space-time coordinates, plus the internal symmetry transformation due to linearity. Naturally, the external symmetry group is Poincar\'e's group, where $a^\mu$ relates to space-time translations, $R^{ij}$ relates to spatial rotations, and $R^{0i}$ corresponds to Lorentz boosts. 

\subsection{Conserved Currents for the Real Scalar Field}
\label{ccrsf}

To obtain the conserved current defined in Eq. (\ref{eqq1}), the 4-vector $\sigma^\mu$ is needed. From Eq. (\ref{eq8}), we obtain
\begin{equation}
\nonumber
\sigma^\mu = \phi\,\partial^{\mu}\tilde\phi \,,
\end{equation}
so that 
\begin{eqnarray}
J^\mu &=& \partial^{\mu}\phi\,a^{\nu}\partial_{\nu}\phi -
a^{\mu}{\cal{L}} +
\partial^{\mu}\phi\,\partial_{\nu}\phi\,R^{\nu}_{\,\alpha}x^\alpha  \nonumber \\
\label{eq22} &-& {\cal{L}}\,R^{\mu}_{\,\nu}x^\nu +
\phi\,\partial^{\mu}\tilde\phi - \tilde\phi\,\partial^{\mu}\phi
\,,
\end{eqnarray}
where ${\cal{L}}$ is given by Eq. (\ref{lsf}). 

It is interesting to examine the conservation laws associated with different symmetries. For time translations, we set 
$a^0 = 1$, together with the remaining parameters and $\tilde\phi$ vanishing. From expression (\ref{eq22}), we get
\begin{eqnarray}
J^0 &=& \frac{1}{2}\,\left[(\partial_{0}\phi)^2 + (\nabla\phi)^2 +
m^{2}\phi^2\right] \,, \nonumber  \\
J^i &=& \partial^{i}\phi\,\partial_{0}\phi \,, \nonumber
\end{eqnarray}
corresponding to the energy conservation law, 
\begin{equation}
\frac{d}{dt} \left\{\frac{1}{2}\,\int \left[(\partial_{0}\phi)^2 + (\nabla\phi)^2 +
m^{2}\phi^2\right] d{\bf r}\right\} = 0 \,. \nonumber
\end{equation}

For space translations, we set $a^\mu = (0,a^{1},a^{2},a^{3})$, together with vanishing remaining parameters and $\tilde\phi$. The conserved currents (one for each component of the translation vector ${\bf a}$) can be expressed as
\begin{eqnarray}
J^0 &=& a^{j}\,\partial^{0}\phi\,\partial_{j}\phi \,,
\label{mom}
\\
J^i &=& a^{j}\,\left[\partial^{i}\phi\,\partial_{j}\phi -
\frac{\delta^{i}_{\,j}}{2}\Bigl((\partial_{0}\phi)^2 - (\nabla\phi)^2 - m^{2}\phi^2\Bigr)\right] \,, \nonumber
\end{eqnarray}
associated with linear momentum conservation, 
\begin{equation}
\frac{d}{dt} \int \partial^0\phi\,\nabla\phi\,d{\bf r} = 0 \,. \nonumber
\end{equation}

Spatial rotations are associated with $R^{i}_{\,j} \neq 0$, which gives 
\begin{eqnarray}
J^0 &=& R^{j}_{\,k}\,x^k\,\partial^{0}\phi\,\partial_{j}\phi\,, \nonumber \\
J^i &=& R^{j}_{\,k}\,x^k\,\left(\partial^{i}\phi\,\partial_{j}\phi - \delta^{i}_{\,j}\,{\cal{L}}\right) \,, \nonumber
\end{eqnarray} 
associated with the angular momentum conservation,
\begin{equation}
\frac{d}{dt} \int \partial^{0}\phi \,\,{\bf r}\times\nabla\phi\,\,d{\bf r} = 0 \,.
\nonumber
\end{equation}

For Lorentz boosts, we take $R^{0}_{\,j} \neq 0$, leaving us with 
\begin{eqnarray}
J^0 &=& R^{0}_{\,j}\left[x^{j}\left(\partial^{0}\phi\,\partial_{0}\phi - {\cal L}\right) - x^0 \,\partial_{0}\phi\,\partial^{j}\phi\right] \,, \nonumber \\ 
J^i &=& R^{0}_{\,j}\left[x^{j}\,\partial^{i}\phi\,\partial_{0}\phi - x^{0}\left(\partial^{i}\phi\,\partial^{j}\phi + {\cal L}\,\delta^{i}_{\,j}\right)\right] \,. \nonumber
\end{eqnarray}
In consequence, 
\begin{equation}
\frac{d}{dt} \int \left[x^0 \,\partial_{0}\phi\,\partial_{i}\phi - x_{i}\,\left(\partial^{0}\phi\,\partial_{0}\phi 
- {\cal L}\right)\right]\,d{\bf r} = 0 \,.
\label{boost}
\end{equation}
Taking into account the momentum density $\pi_i = \partial_{0}\phi\,\partial_{i}\phi$ which follows from Eq. (\ref{mom}) and the energy density ${\cal H} = \partial^{0}\phi\,
\partial_{0}\phi - {\cal L}$, Eq. (\ref{boost}) reads 
\begin{equation}
\frac{d}{dt}\,\int\left(x^{0}\,\boldsymbol{\pi} - {\cal H}\,{\bf r}\right)\,d{\bf r} = 0 \,, \nonumber
\end{equation}
showing, in a more transparent way, that the linear momentum relative to the center of mass is constant.  Notice that the Noether momentum density, a quantity arising from space translation symmetries, in this case is not the same as the canonical momentum density $\partial{\cal L}/\partial(\partial_{0}\phi) = \partial^{0}\phi$. 
Similarly, for a free relativistic particle with momentum ${\bf p}$ and energy $H$, one has $(d/dt)({\bf p}\,t - H\,{\bf r}) = 0$.

Finally, for the internal symmetry, only $\tilde\phi$ is non-vanishing. In this case, 
\begin{equation}
J^\mu = \phi\,\partial^{\mu}\tilde\phi - \tilde\phi\,\partial^{\mu}\phi \quad \Rightarrow \quad \frac{d}{dt}\int \left(\phi\,\partial^{0}\tilde\phi - \tilde\phi\,\partial^{0}\phi\right)\,d{\bf r} = 0 \,. \nonumber
\end{equation}
This conserved current is analogous to the constant wronskian of two par\-ti\-cu\-lar solutions $y_{1,2}(t)$ for a
linear harmonic oscillator equation, namely,
\begin{equation}
\frac{d^2 y_{1,2}}{dt^2} + y_{1,2} = 0 \quad \Rightarrow \quad \frac{d}{dt}\left(y_1 \frac{dy_2}{dt} - y_2 \frac{dy_1}{dt}\right) = 0 \,. \nonumber
\end{equation}

Recapitulating, we have derived, in a systematic way, the Noether point symmetries group for the real scalar field, obtaining the Poincar\'e group (external symmetries) plus an infinite dimensional internal symmetry group, reflecting the linearity of the Klein-Gordon equation. It can be mentioned that such internal Noether symmetry transformation 
 is not recognized in the literature, as far as we know.  

\section{Complex Scalar Field}
\label{csf}

In this Section, we consider the case of the complex scalar field, so that a global gauge symmetry is expected. Let us verify this, proceeding in a systematic manner, as done in the case of the real scalar field. 

\subsection{Noether Symmetries for the Complex Scalar Field}

The Lagrangian density for the complex scalar field is 
\begin{equation}
{\cal L} = \partial^{\mu}\phi^{*}\,\partial_{\mu}\phi -
m^{2}\,\phi^{*}\,\phi \,,
\label{ldcsf}
\end{equation}
implying two separate equations for the independent fields $\phi, \phi^*$,
\begin{eqnarray}
\partial^{\mu}\partial_{\mu}\phi + m^{2}\phi &=& 0 \,, \nonumber \\  \partial^{\mu}\partial_{\mu}\phi^* + m^{2}\phi^*
&=& 0 \, \nonumber
\end{eqnarray}
Due to the linearity of the equations, it is reasonable to suppose the existence of internal Noether symmetries corresponding to the addition of particular solutions. 

Consider the infinitesimal point transformations, 
\begin{eqnarray}
x^\mu &\rightarrow& x^{\mu} +
\varepsilon\,\eta^{\mu}(\phi,\phi^{*},x) \,, \nonumber
\\ \phi &\rightarrow& \phi + \varepsilon\,\psi(\phi,\phi^{*},x)
\,, \nonumber \\  \phi^* &\rightarrow& \phi^* +
\varepsilon\,\psi^{*}(\phi,\phi^{*},x)\,, \nonumber
\end{eqnarray}
where $\varepsilon$ is an infinitesimal parameter, and where 
$\eta^{\mu}(\phi,\phi^{*},x)$, $\psi(\phi,\phi^{*},x)$, $\psi^{*}(\phi,\phi^{*},x)$ 
are smooth functions to be determined and not depending on field derivatives. The Noether invariance condition (\ref{eq1}), generalized to more than a single field, yields 
\begin{eqnarray}
 \frac{\partial{\cal{L}}}{\partial\phi}\,\psi +
\frac{\partial{\cal{L}}}{\partial\phi^*}\,\psi^* + \frac{\partial
{\cal{L}}}{\partial(\partial_{\mu}\phi)}\,(d_{\mu}\psi -
\partial_{\nu}\phi\,d_{\mu}\eta^{\nu}) &+& \nonumber \\ \label{equa1} +
\frac{\partial
{\cal{L}}}{\partial(\partial_{\mu}\phi^*)}\,(d_{\mu}\psi^*
-
\partial_{\nu}\phi^{*}\,d_{\mu}\eta^{\nu}) +
\partial_\mu{\cal{L}}\,\eta^\mu +
{\cal{L}}\,d_{\mu}\eta^\mu &=& d^{\mu}\sigma_\mu \,,
\nonumber
\end{eqnarray}
where $\sigma_\mu = \sigma_{\mu}(\phi,\phi^{*},x)$ is, at this stage, an arbitrary 4-vector.  The corresponding conserved current reads
\begin{equation}
\label{eq300} J^\mu = \theta^{\mu}_{\,\nu}\,\eta^\nu -
\frac{\partial{\cal L}}{\partial(\partial_{\mu}\phi)}\,\psi -
\frac{\partial{\cal L}}{\partial(\partial_{\mu}\phi^{*})}\,\psi^* +
\sigma^\mu \,,
\end{equation}
where the energy-momentum tensor
\begin{equation}
\theta^{\mu}_{\,\nu} = \frac{\partial{\cal
L}}{\partial(\partial_{\mu}\phi)}\,\partial_{\nu}\phi +
\frac{\partial{\cal
L}}{\partial(\partial_{\mu}\phi^*)}\,\partial_{\nu}\phi^* -
\delta^{\mu}_{\,\nu}\,{\cal L} \,. \nonumber
\end{equation}
was used. 

Inserting ${\cal{L}}$ from Eq. (\ref{ldcsf}) into the symmetry condition (\ref{eq1}),
it results a polynomial on the derivatives of $\phi, \phi^*$, to be identically set to zero. The coefficient of each monomial (term with equal power of the derivatives of the fields) should then vanish. For brevity, we will not show 
the full calculations, which are analogous to the case of the real scalar field. 

The terms of degree three imply 
\begin{equation}
\eta^{\mu} = \eta^{\mu}(x) \,,
\nonumber
\end{equation}
so that the external transformations are field-independent. 

The terms of degree three and two imply 
\begin{eqnarray}
\label{eq2} \frac{\partial\psi}{\partial\phi^*} =
\frac{\partial\psi^*}{\partial\phi} &=& 0 \,,\\ \label{eq3}
\delta_{\mu\nu}\Bigl(\frac{\partial\psi}{\partial\phi} +
\frac{\partial\psi^*}{\partial\phi^*} &+&
\partial_{\alpha}\eta^\alpha\Bigr) - \partial_{\mu}\eta_\nu -
\partial_{\nu}\eta_\mu = 0 \,.
\end{eqnarray}
A tedious analysis shows that Eqs. (\ref{eq2}) and (\ref{eq3}) satisfy the appropriate Cauchy conditions if and only if 
\begin{eqnarray}
\label{x1}
\partial_{1}\eta_1 &=& \partial_{2}\eta_2 = \partial_{3}\eta_3 = -
\partial_{0}\eta_0  \,, \nonumber \\
\partial_{\mu}\eta_\nu &+& \partial_{\nu}\eta_\mu = 0 \,,\quad \mu \neq
\nu \,. \nonumber
\label{xx1}
\end{eqnarray}
In this case, the solution for Eqs. (\ref{eq2})-(\ref{eq3}) is 
\begin{eqnarray}
\psi &=& - \phi\,\partial_{0}\eta_0 - i\,\lambda(x)\phi +
\tilde\phi(x) \,, \label{p1} \\ \psi^* &=& - \phi^{*}\,\partial_{0}\eta_0 +
i\,\lambda(x)\phi^* + \tilde\phi^{*}(x) \,, \label{p2}
\end{eqnarray}
where $\lambda(x)$, $\tilde\phi(x)$ and $\tilde\phi^{*}(x)$ are arbitrary functions not depending on the fields. For consistency, 
\begin{equation}
\label{x2}
\lambda(x) = \lambda^{*}(x) \,.
\end{equation}
The terms depending on $\lambda(x)$ are associated with local gauge transformations. 

The invariance condition, regarding terms of first order in the derivatives of the fields, gives 
\begin{eqnarray}
\frac{\partial\sigma_\mu}{\partial\phi} &=&
\partial_\mu\psi^* \,, \nonumber \\
\frac{\partial\sigma_\mu}{\partial\phi^*} &=&
\partial_\mu\psi \,. \nonumber
\end{eqnarray}
Taking into account $\psi$, $\psi^*$ from Eqs. (\ref{p1}) and (\ref{p2}), it can be verified that the Cauchy condition
\begin{equation}
\frac{\partial^{2}\sigma_\mu}{\partial\phi\partial\phi^*} =
\frac{\partial^{2}\sigma_\mu}{\partial\phi^{*}\partial\phi} \label{cau}
\end{equation}
yields 
\begin{equation}
\partial_{\mu}\lambda(x) = 0 \,. 
\nonumber
\end{equation}
In other words, $\lambda$ does not depend on $x$ and gauge transformations are global. Besides, in this case we can solve for 
\begin{equation}
\label{eq500} \sigma_\mu = -
\phi^{*}\phi\,\partial_{0}\partial_{\mu}\eta_0 +
\phi^{*}\partial_{\mu}\tilde\phi + \phi\,\partial_{\mu}\tilde\phi^*
+ \tilde\sigma_{\mu}(x) \,,
\end{equation}
where $\tilde\sigma_{\mu}(x)$ is an arbitrary 4-vector, depending on $x$ only. 

Finally, the term not containing field derivatives in the invariance condition (\ref{eq1}) gives
\begin{eqnarray}
\phi^{*}\phi\left(\partial^{\mu}\partial_{\mu}\partial_{0}\eta_0 -
2m^{2}\partial_{0}\eta_0\right) -
\phi^{*}\left(\partial^{\mu}\partial_{\mu}\tilde\phi +
m^{2}\tilde\phi\right) \nonumber \\ -
\phi\left(\partial^{\mu}\partial_{\mu}\tilde\phi^* +
m^{2}\tilde\phi^*\right) - \partial^{\mu}\tilde\sigma_\mu = 0 \,. \nonumber
\end{eqnarray}
The terms proportional to $\phi^{*}\phi$, $\phi^*$, $\phi$ and the remaining contribution should identically vanish. Hence,
\begin{eqnarray}
\label{e4}
\partial_{0}(\partial^{\mu}\partial_{\mu}\eta_0 - 2m^{2}\eta_0) &=& 0 \,,\\
\label{eq5} \partial^{\mu}\partial_{\mu}\tilde\phi + m^{2}\tilde\phi &=& 0 \,,\\
\label{eq6} \partial^{\mu}\partial_{\mu}\tilde\phi^* + m^{2}\tilde\phi^* &=& 0 \,,\\
\label{eq7}
\partial^{\mu}\tilde\sigma_\mu &=& 0 \,.
\end{eqnarray}
Equations (\ref{eq5}) and (\ref{eq6}) show that $\tilde\phi$
and $\tilde\phi^*$ are particular solutions of the Klein-Gordon equation, while Eq. (\ref{eq7})
shows that 
\begin{equation}
\tilde\sigma_\mu = 0 \,,
\nonumber
\end{equation}
without loss of generality. Finally, Eq. (\ref{e4}) reveals that 
\begin{equation}
\label{ult}
\partial^{\mu}\partial_{\mu}\eta_0 - 2m^{2}\eta_0 = - 2m^{2}\tilde\eta_{0}({\bf r}) \,,
\end{equation}
where $\tilde\eta_{0}({\bf r})$ is an arbitrary time-independent function. Equation (\ref{ult}) coincides with 
Eq. (\ref{eqq300}). In this way, we realize that there is no need to repeat the previous calculations, with the conclusion that the external Noether symmetries are given by the Poincar\'e group. Therefore, 
\begin{eqnarray}
\eta^\mu &=& = a^\mu + R^{\mu}_{\,\nu}x^\nu  \,, \nonumber \\ \psi &=& -
i\lambda\,\phi + \tilde\phi(x) \,, \nonumber \\ \psi^* &=& i\lambda\,\phi^* +
\tilde\phi^{*}(x) \,, \nonumber
\end{eqnarray}
where $a^\mu$ is a constant 4-vector, $R^{\mu}_{\,\nu}$ is a second-rank antisymmetric tensor,  $\lambda$ is a real constant and 
$\tilde\phi, \tilde\phi^*$ are particular solutions of the Klein-Gordon equation. To conclude, besides the Poincar\'e group, one has the internal symmetries, composed by a global gauge transformation and a symmetry due to the linearity of the model.

\subsection{Conserved Currents for the Complex Scalar Field}

To compute the conserved currents, there is the need of the 4-vector 
$\sigma_\mu$, obtained from Eq. (\ref{eq500}),
\begin{equation}
\sigma^\mu = \phi^{*}\partial^{\mu}\tilde\phi +
\phi\,\partial^{\mu}\tilde\phi^* \,. \nonumber
\end{equation}
Inserting this 4-vector into $J^\mu$ in Eq. 
(\ref{eq300}), we get a somewhat long expression. In comparison to the case of the real scalar field, the distinctive feature comes from the global gauge symmetry. Setting $\lambda = 1$ and the remaining contributions to zero, we derive the Noether's current, 
\begin{equation}
J^{\mu}_{(\lambda)} = i\,(\phi\,\partial^{\mu}\phi^* -
\phi^{*}\partial^{\mu}\phi) \,, \nonumber 
\end{equation}
corresponding to global electric charge conservation, 
\begin{equation}
\frac{d}{dt}\left[\,i\,\int (\phi\,\partial^{0}\phi^* -
\phi^{*}\partial^{0}\phi)\,d{\bf r}\right] = 0 \,. \nonumber
\end{equation}
The remaining Noether currents are analogous to those derived in Section \ref{ccrsf}.

\section{Vacuum Electromagnetic Field}
\label{ef}

\subsection{Noether Symmetries for the Electromagnetic Field in Vacuum}

Following our schedule, the Noether point symmetries for the vacuum electromagnetic field will be studied, without {\it ad hoc} claims. The Lagrangian density is
\begin{equation}
{\cal L} = - \frac{1}{4}\,F^{\mu\nu}\,F_{\mu\nu} \,,
\label{vl}
\end{equation}
where
\begin{equation}
F_{\mu\nu} = \partial_{\mu}A_\nu - \partial_{\nu}A_\mu \nonumber
\end{equation}
is the electromagnetic tensor, while $A_\mu = (A_{0},{\bf A})$ denotes the electromagnetic 4-potential. The 
Euler-Lagrange equations are the vacuum Maxwell's equations, 
\begin{equation}
\partial^{\nu}\partial_{\nu}\,A^\mu - \partial^{\mu}\partial_{\nu}A^\nu = 0 \,. 
\nonumber
\end{equation}
We can anticipate the existence of internal Noether symmetries, associated with the linearity of the equations. 

Suppose the infinitesimal point transformations, 
\begin{eqnarray}
x^\mu &\rightarrow& x^{\mu} + \varepsilon\,\eta^{\mu}(A,x) \,, \nonumber
\\ A^\mu &\rightarrow& A^\mu + \varepsilon\,\Gamma^{\mu}(A,x) \nonumber
\,,
\end{eqnarray}
where $A$ is a shorthand for the 4-potential, $\varepsilon$ an infinitesimal parameter, and $\eta^{\mu}(A,x)$, $\Gamma^{\mu}(A,x)$ 
functions to be determined, not depending on field derivatives. 
Since the Lagrangian density contains only field derivatives, the Noether symmetry condition simplifies to  
\begin{equation}
\frac{\partial{\cal
L}}{\partial(\partial_{\mu}A_\nu)}\,\left(d_{\mu}\Gamma_\nu -
\partial_{\alpha}A_{\nu}\,d_{\mu}\eta^\alpha\right) +
{\cal{L}}\,d_{\mu}\eta^\mu =
d^{\mu}\sigma_\mu \,,
\label{cn}
\end{equation}
where, at this stage,  $\sigma_\mu = \sigma_{\mu}(A,x)$ is an arbitrary 4-vector with the indicated functional dependence.

The procedure must be clear now. Inserting ${\cal L}$ from Eq. (\ref{vl}) into Eq. (\ref{cn}) and considering field derivatives of third degree, it follows that 
\begin{equation}
\eta^{\mu} = \eta^{\mu}(x) \,. \nonumber
\end{equation}
Once again, the space-time transformation rules do not depend on the fields. 

After several elementary calculations, the second-order field derivative terms give 
%
\begin{eqnarray}
\label{f1} \Gamma^\mu =  - \partial^{\mu}\eta_{\nu}\,\,A^\nu &+&
\tilde{A^{\mu}}(x) \,,\\ \label{ff1}
\partial_{\mu}\eta_\nu + \partial_{\nu}\eta_\mu &=& 0 \,,
\end{eqnarray}
where the 4-vector $\tilde{A^\mu}$ is field-independent. 
As seen before, the Poincar\'e group is obtained from Eq. (\ref{ff1}).

The first-order in the field derivative terms give 
\begin{equation}
\partial_\mu\Gamma_\nu -
\partial_\nu\Gamma_\mu =
\frac{\partial\sigma_\nu}{\partial A^\mu} \,, \nonumber
\end{equation}
or, taking into account Eq. (\ref{f1}),
\begin{equation}
\partial_\mu{\tilde A}_\nu -
\partial_\nu{\tilde A}_\mu =
\frac{\partial\sigma_\nu}{\partial A^\mu} \,, \nonumber
\end{equation}
with the solution 
\begin{equation}
\sigma_\mu = (\partial_{\nu}\tilde{A_\mu} -
\partial_{\mu}\tilde{A_{\nu}})\,A^\nu + \tilde\sigma_{\mu}(x) \nonumber
\,,
\end{equation}
where $\tilde\sigma_{\mu}(x)$ is an arbitrary field-independent 4-vector.

From the remaining term in the invariance condition, we get 
\begin{eqnarray}
\label{e2} \partial^{\mu}\tilde\sigma_\mu &=& 0 \,, \\ \label{e3}
\partial^{\nu}\partial_{\nu}\tilde{A^\mu}
-
\partial^{\mu}(\partial_{\nu}\tilde{A^\nu}) &=& 0 \,.
\end{eqnarray}
While Eq. (\ref{e2}) shows that $\tilde\sigma_\mu$ is superfluous, Eq. 
(\ref{e3}) implies that $\tilde{A^\mu}$ is a particular solution for vacuum Maxwell's equations. 

Summing up results, and using Eq. (\ref{f1}), we derive the following symmetry transformation functions, 
\begin{eqnarray}
\eta^\mu &=& a^\mu + R^{\mu}_{\,\nu}x^\nu \,,\quad 
R_{\mu\nu} + R_{\nu\mu} = 0 \,, \nonumber \\ \label{eq4} \Gamma^\mu &=& \tilde{A^\mu}(x) + 
R^{\mu}_{\,\nu}A^\nu  \,, \nonumber
\end{eqnarray}
where $\tilde{A^\mu}(x)$ solves Maxwell's vacuum equations. 

A notorious particular vacuum solution is specified by local gauge transformations, 
\begin{equation}
\tilde{A^\mu} = \partial^{\mu}\lambda(x) \,, \nonumber
\end{equation}
where $\lambda(x)$ is an arbitrary space-time function. In this context there is a link between  
the linearity of vacuum Maxwell's equations and local gauge transformations.

\subsection{Conserved Currents for the Electromagnetic Field in Vacuum}

The conserved current in all generality reads
\begin{eqnarray}
J^\mu &=& a^\nu\,\left(- F^{\mu\alpha}\,\partial_{\nu}A_\alpha + \frac{1}{4}\,\delta^{\mu}_{\,\nu}\,F^{\alpha\beta}F_{\alpha\beta}\right) \nonumber \\
&+& R^{\alpha}_{\,\beta}\left(-x^\beta\,F^{\mu\nu}\,\partial_\alpha A_\nu + \frac{1}{4}\,x^\beta \delta^{\mu}_{\,\alpha}\,F^{\nu\gamma}\,F_{\nu\gamma} + A^\beta\,F^{\mu}_{\,\,\alpha}\right) \nonumber \\
&+&
F^{\mu\nu}\tilde{A}_\nu - \tilde{F}^{\mu\nu}\,A_\nu \,, \label{eff}
\end{eqnarray}
where it was defined $\tilde{F}^{\mu\nu} = \partial^{\mu}\tilde{A}^\nu - \partial^{\nu}\tilde{A}^\mu$. 

The usual symmetries (space-time translations and spatial rotations) are fairly well discussed in the literature, including additional steps such as the symmetrization of the energy-momentum tensor \cite{Kleinert}. For instance, in terms of the electric field $E_i = - F_{0i}$ and the magnetic field $B_i = - (1/2)\,\varepsilon_{ijk}\,F^{jk}$, where $\varepsilon_{ijk}$ is the 3-dimensional Kronecker symbol, for time-translations (only $a^0 = 1$ is non-zero) one finds 
\begin{equation}
J^0 = \frac{1}{2}\,(E^2 + B^2) + \nabla\cdot(A_0\,{\bf E}) - A_0\,\nabla\cdot{\bf E} \,. \nonumber
\end{equation}
The third term is a surface term and so does not contribute to the associated conserved quantity. The last term vanishes since $\nabla\cdot{\bf E} = 0$ in vacuum. The only effective contribution is 
\begin{equation}
\int\,J^{0}\,d{\bf r} = \frac{1}{2}\,\int\,(E^2 + B^2)\,d{\bf r}  \,, \nonumber 
\end{equation}
which is the well-known electromagnetic energy density. 

Actually the conserved current (\ref{eff}) can be put into a more traditional and gauge-invariant form, as follows. One has 
\begin{eqnarray}
J^\mu &=& a^\nu\,\left(- F^{\mu\alpha}\,F_{\nu\alpha} + \frac{1}{4}\,\delta^{\mu}_{\,\nu}\,F^{\alpha\beta}F_{\alpha\beta} - \partial_\alpha(F^{\mu\alpha}\,A_\nu) + (\partial_\alpha F^{\mu\alpha})\,A_\nu\right) \nonumber \\
&+& \frac{1}{2}\,R^{\alpha\beta}\Bigl[x^\beta\,\left(- F^{\mu\nu}\,F_{\alpha\nu} + \frac{1}{4}\,\delta^{\mu}_{\,\alpha}\,F^{\nu\gamma}\,F_{\nu\gamma}\right) - x^\alpha\,\left(- F^{\mu\nu}\,F_{\beta\nu} + \frac{1}{4}\,\delta^{\mu}_{\,\beta}\,F^{\nu\gamma}\,F_{\nu\gamma}\right)
\nonumber \\ &+& \partial_\nu\left(x_\alpha F^{\mu\nu}\,A_\beta - x_\beta F^{\mu\nu}\,A_\alpha\right) - (x_\alpha\,A_\beta - x_\beta\,A_\alpha)\,\partial_\nu\,F^{\mu\nu} \Bigr] \nonumber \\
&+&
F^{\mu\nu}\tilde{A}_\nu - \tilde{F}^{\mu\nu}\,A_\nu \,, \label{efff}
\end{eqnarray}
Most terms in (\ref{efff}) are total divergence terms, integrating to zero, or vanish due to Maxwell's equations in vacuum. Therefore essentially we have the well-known \cite{Hill, Kleinert} results, 
\begin{equation}
\label{rec}
J^\mu = a^\nu P^{\mu}_{\,\nu} + \frac{1}{2}\,R^{\alpha\beta}\,M_{\beta\alpha}^\mu + F^{\mu\nu}\tilde{A}_\nu - \tilde{F}^{\mu\nu}\,A_\nu \,,
\end{equation}
where 
\begin{equation}
P^{\mu}_{\,\nu} = - F^{\mu\alpha}\,F_{\nu\alpha} + \frac{1}{4}\,\delta^{\mu}_{\,\nu}\,F^{\alpha\beta}F_{\alpha\beta} \nonumber
\end{equation}
is associated to momentum conservation and 
\begin{equation}
M_{\alpha\beta}^\mu = x_\alpha\,P^{\mu}_{\,\beta} - x_\beta\,P^{\mu}_{\,\alpha} \nonumber
\end{equation}
is related to angular momentum conservation. As can be checked, $\partial_\mu\,J^\mu = 0$ is maintained in the reshaped form (\ref{rec}).

The internal symmetry due to linearity is basically ignored in the usual treatments. As apparent from Eq. (\ref{eff}), one has the conservation law
\begin{equation}
\nonumber
\frac{d}{dt}\,\int (F^{0\mu}\tilde{A}_\mu - \tilde{F}^{0\mu}\,A_\mu)\,d{\bf r} = 0 \,,
\end{equation}
where both $A_\mu, \tilde{A}_\mu$ solve Maxwell's equations. 

The case of local gauge symmetries with $\tilde{A}_\mu = \partial_{\mu}\lambda$  yields $\partial_{\mu}J^{\mu}_{(\lambda)} = 0$, where 
\begin{equation}
J^{\mu}_{(\lambda)} = 
F^{\mu\nu}\partial_{\nu}\lambda = \partial_\nu(F^{\mu\nu}\lambda) - \lambda\,\partial_\nu\,F^{\mu\nu}\,. \label{rhs}
\end{equation}
The subscript $\lambda$ refers to the particular gauge function employed. At least for $\lambda$ bounded at infinity, the first term in the right-hand side of Eq. (\ref{rhs}) contribute a vanishing surface term, while the last term vanishes due to Maxwell's equations in vacuum. Hence in this case ($\tilde{A}_\mu = \partial_\mu\lambda$ for some function $\lambda$) the related conserved charge vanishes. Otherwise it can happens that the conservation law yields a non-trivial result. 

\section{Coupled Complex Scalar and Electromagnetic Fields}
\label{csem}

\subsection{Noether Symmetries for the Coupled Complex Scalar and Electromagnetic Fields}

The Lagrangian density follows from the minimal coupling assumption and is given by 
\begin{equation}
{\cal L} = (\partial^{\mu}\phi^* -
i\,e\,A^{\mu}\,\phi^{*})(\partial_{\mu}\phi + i\,e\,A_{\mu}\,\phi) -
m^{2}\phi^{*}\phi - \frac{1}{4}\,F^{\mu\nu}F_{\mu\nu} \,, \label{laga}
\end{equation}
where $e$ is the particle's charge and $F^{\mu\nu}$ the electromagnetic tensor, as before. 
The Euler-Lagrange equations are 
\begin{eqnarray}
D^{\mu}D_{\mu}\phi + m^{2}\phi &=& 0 \,, \nonumber \\ (D^{\mu}D_{\mu}\phi)^* +
m^{2}\phi^* &=& 0 \,, \nonumber \\ 
\partial_\nu F^{\mu\nu} &=& i\,e\, \Bigl(\phi\,(D^{\mu}\phi)^* -
\phi^*\,(D^{\mu}\phi)\Bigr) \nonumber
\,,
\end{eqnarray}
where
\begin{eqnarray}
D^{\mu}\phi &=& (\partial^\mu + i\,e\,A^{\mu})\,\phi \,, \nonumber \\
(D^{\mu}\phi)^* &=& (\partial^\mu - i\,e\,A^{\mu})\,\phi^* \nonumber
\end{eqnarray}
denote covariant derivatives. Due to presence of matter and the corresponding nonlinearity, we expect a broken internal symmetry, previously corresponding to the linear superposition law. 

Consider infinitesimal point transformations of the form
\begin{eqnarray}
x^\mu &\rightarrow& x^{\mu} +
\varepsilon\,\eta^{\mu}(A,\phi,\phi^{*},x) \,, \nonumber
\\
A^\mu &\rightarrow& A^\mu + \varepsilon\,\Gamma(A,\phi,\phi^{*},x)
\,, \nonumber
\\ \phi &\rightarrow& \phi + \varepsilon\,\psi(A,\phi,\phi^{*},x)
\,, \nonumber \\ \phi^* &\rightarrow& \phi^* +
\varepsilon\,\psi^{*}(A,\phi,\phi^{*},x) \,. \nonumber
\end{eqnarray}

The Noether symmetry condition is 
\begin{eqnarray}
\frac{\partial{\cal L}}{\partial\,A^\mu}\,\Gamma^\mu
&+& \frac{\partial{\cal L}}{\partial\phi}\,\psi +
\frac{\partial{\cal L}}{\partial\phi^*}\,\psi^*
+
\frac{\partial{\cal L}}{\partial
(\partial_{\mu}A_\nu)}\,(d_{\mu}\Gamma_\nu
-
\partial_{\rho}A_{\nu}\,d_{\mu}\eta^\rho)
+ \nonumber \\ &+&
\frac{\partial{\cal{L}}}{\partial(\partial_{\mu}\phi)}\,(d_{\mu}\psi
- \partial_{\nu}\phi\,d_{\mu}\eta^{\nu}) +
\frac{\partial{\cal{L}}}{\partial(\partial_{\mu}\phi^*)}\,
(d_{\mu}\psi^*
- \partial_{\nu}\phi^{*}\,d_{\mu}\eta^{\nu}) + \nonumber
\\ &+& {\cal{L}}\,d_{\mu}\eta^\mu =
d^{\mu}\sigma_\mu \,, \nonumber
\end{eqnarray}
where $\sigma_\mu = \sigma_{\mu}(A,\phi,\phi^{*},x)$ is an arbitrary 4-vector. 

From the cubic in the derivatives terms, it can be shown that 
\begin{equation}
\eta^{\mu} = \eta^{\mu}(x) \,. \nonumber
\end{equation}
The quadratic terms give 
\begin{eqnarray}
\label{e1} \Gamma^\mu \nonumber &=& - \partial^{\mu}\eta_{\nu}\,\,A^\nu +
\tilde{A^{\mu}}(x) \,,\\ \psi &=& - i\,e\,\lambda(x)\,\phi +
\tilde\phi(x) \,, \nonumber \\ \psi^* &=& i\,e\,\lambda(x)\,\phi^* +
\tilde\phi^{*}(x) \,, \nonumber \\ \label{eqqq1}
\partial_{\mu}\eta_\nu &+& \partial_{\nu}\eta_\mu = 0 \,,
\end{eqnarray}
where $\tilde{A^\mu}$, $\tilde\phi$, $\tilde\phi^*$ and 
$\lambda$ are field independent. The factor $e$ was included for convenience.  Moreover, $\lambda$ is a real function. Proceeding as in the vacuum case, from Eq. (\ref{eqqq1}) it follows that the external symmetries are associated with the Poincar\'e group. 

The linear in the derivative terms imply 
\begin{eqnarray}
\label{eq30}
\frac{\partial\sigma_\nu}{\partial\,A^\mu} &=&
\partial_\mu\tilde{A_\nu} -
\partial_\nu\tilde{A_\mu} \,,\\
\label{eq31}
\frac{\partial\sigma_\mu}{\partial\phi} &=& -
i\,e\,A_{\mu}\,\tilde\phi^* - i\,e\,\phi^{*}\,\tilde{A_\mu} +
i\,e\,(\partial_{\mu}\lambda)\,\phi^* + \partial_{\mu}\tilde\phi^* \,,\\
\label{eq32}
\frac{\partial\sigma_\mu}{\partial\phi^*} &=&
i\,e\,A_{\mu}\,\tilde\phi + i\,e\,\phi\,\tilde{A_\mu} -
i\,e\,(\partial_{\mu}\lambda)\,\phi + \partial_{\mu}\tilde\phi \,.
\end{eqnarray}
The integrability condition
\begin{equation} 
\frac{\partial^{2}\sigma_\mu}{\partial\phi\,\partial\phi^*} =
\frac{\partial^{2}\sigma_\mu}{\partial\phi^{*}\,\partial\phi} \nonumber
\end{equation}
implies
\begin{equation}
\label{eq40} \tilde{A_\mu} = \,\partial_{\mu}\lambda
\,.
\end{equation}
From Eq. (\ref{eq40}) it can be seen that in the presence of matter only (local) gauge symmetries are allowed.

Once Eq. (\ref{eq40}) is satisfied, it is possible to solve Eqs. (\ref{eq30})-(\ref{eq32}), with the result 
\begin{eqnarray}
\tilde\phi &=& \tilde\phi^* = 0 \,,\nonumber \\ \sigma_\mu &=&
\tilde\sigma_{\mu}(x) \,. \nonumber
\end{eqnarray}
As verified, the symmetries due to the linearity are completely eliminated. 

The remaining terms in the invariance condition give 
\begin{equation}
\partial^{\mu}\tilde\sigma_\mu = 0 \,, \nonumber
\end{equation}
so that there is no loss of generality to define
\begin{equation}
\tilde\sigma_\mu = 0 \,.
\nonumber
\end{equation}

To conclude, the external symmetries are composed by the Poincar\'e transformations, while the internal symmetries are specified by 
\begin{eqnarray}
\Gamma^\mu &=& R^{\mu}_{\,\nu}A^\nu +
\,\partial^{\mu}\lambda(x) \,, \nonumber \\
\psi &=& - i\,e\,\lambda(x)\,\phi \,, \nonumber \\ \psi^* &=& i\,e\,\lambda(x)\,\phi^* \,. \nonumber
\end{eqnarray}

\subsection{Conserved Currents for the Coupled Complex Scalar and Electromagnetic Fields}

The general conserved current is
\begin{eqnarray}
J^{\mu} &=& a^{\nu}\,\Bigl(\partial_\nu\phi\,(D^\mu\phi)^* + \partial_\nu\phi^*\,(D^\mu\phi) - F^{\mu\alpha}\,\partial_\nu A_\alpha - \delta^{\mu}_{\,\nu}{\cal L} \Bigr)  \nonumber \\
&+& R^{\alpha}_{\,\beta} \Bigl[x^\beta \Bigl(\partial_\alpha\phi\,(D^\mu\phi)^* + \partial_\alpha\phi^*\,(D^\mu\phi) - F^{\mu\nu}\,\partial_\alpha A_\nu - \delta^{\mu}_{\,\alpha}{\cal L}\Bigr) + A^\beta\,F^{\mu}_{\alpha}\Bigr] \nonumber \\
&+& F^{\mu\nu}\,\partial_\nu\lambda + i\,e\,\lambda\,\phi\,(D^\mu\phi)^* - i\,e\,\lambda\,\phi^*\,(D^\mu\phi) \,. \label{pen} \nonumber
\end{eqnarray}

The usual conservation laws (energy, linear momentum, angular momentum) are verified. 
Besides, we have the gauge symmetry, with the conserved current 
\begin{equation}
J^{\mu}_{(\lambda)} = F^{\mu\nu}\,\partial_\nu\lambda + i\,e\,\lambda\,\phi\,(D^\mu\phi)^* - i\,e\,\lambda\,\phi^*\,(D^\mu\phi) \,.
\nonumber
\end{equation}
In particular, 
\begin{equation}
J^{0}_{(\lambda)} = \partial_{i}(F^{0i}\lambda) - \lambda\,\left[\partial_{i}F^{0i} + i\,e\,\phi^{*}\,(D^\mu\phi) - i\,e\,\phi\,(D^\mu\phi)^*\right] = \partial_{i}(F^{0i}\lambda) \,,
\nonumber
\end{equation}
the last equality coming from Maxwell's equations. 
Therefore, $dQ_{(\lambda)}/dt = 0$, where 
\begin{equation}
Q_{(\lambda)} = \int\,J^{0}_{(\lambda)}\,d{\bf r} = - \int\,\nabla\cdot(\lambda\,{\bf E})\,d{\bf r} \,,
\nonumber
\end{equation}
which is the electric charge for $\lambda = - 1$, where $E_i = - F_{0i}$ is the electric field. 

\section{Charged Scalar Particle under External Electro\-magnetic Fields}

The Lagrangian density is the same as in Eq. (\ref{laga}), omitting the free electromagnetic field contribution, but retaining the interaction terms, 
\begin{equation}
{\cal L} = (\partial^{\mu}\phi^* -
i\,e\,A^{\mu}\,\phi^{*})(\partial_{\mu}\phi + i\,e\,A_{\mu}\,\phi) -
m^{2}\phi^{*}\phi \,. \label{lagar} \nonumber
\end{equation}
The interpretation is that $A_\mu$ is a given external field, acting on a test charge. Such problem has implications in laser-plasma interactions in the quantum relativistic regime \cite{Marklund}. 

The Noether symmetry condition is given by Eq. (\ref{eq1}), since $A_\mu$ is not subject to any transformation. The treatment of the third and second-order in the field derivative terms yields the results already shown in Eqs. (\ref{x1})-(\ref{x2}), with the replacement $\lambda \rightarrow e\,\lambda$ since now there is the test charge $e$. 

From the first and zeroth order in the field derivatives terms, the following equations are derived, 
\begin{eqnarray}
\frac{\partial\sigma_\mu}{\partial\phi} &=& \Bigl(i\,e\,A_\nu\,\partial^{\nu}\eta_\mu - \partial_{\mu}\partial_{0}\eta_0 + i\,e\,\partial_\mu\lambda - i\,e\,\eta_\nu\partial^{\nu}A_\mu - 2\,i\,e\,(\partial_0\eta_0)\,A_\mu\Bigr)\,\phi^* \nonumber \\ &+& \partial_\mu\tilde{\phi}^* - i\,e\,A_\mu\,\tilde{\phi}^* \,, \label{fu1}\\
\frac{\partial\sigma_\mu}{\partial\phi^*} &=& \Bigl(- i\,e\,A_\nu\,\partial^{\nu}\eta_\mu - \partial_{\mu}\partial_{0}\eta_0 - i\,e\,\partial_\mu\lambda + i\,e\,\eta_\nu\partial^{\nu}A_\mu + 2\,i\,e\,(\partial_0\eta_0)\,A_\mu\Bigr)\,\phi \nonumber \\ &+& \partial_\mu\tilde{\phi} + i\,e\,A_\mu\,\tilde{\phi} \,, \label{fu2}\\
\partial^\mu\sigma_\mu &=& 2\,\Bigl(-\,e^2\,A^\mu\partial_\mu\lambda + \,e^2\eta_\nu\,(\partial^\nu A_\mu)\,A^\mu + (\partial_{0}\eta_0)\,(e^2\,A^\mu A_\mu - m^2)\Bigr)\,\phi^*\,\phi \nonumber \\
&+& \Bigl(i\,e\,A^\mu\partial_\mu\tilde{\phi}^* + (e^2\,A^\mu A_\mu - m^2)\,\tilde{\phi}^*\Bigr)\,\phi \nonumber \\ &+& \Bigl(- i\,e\,A^\mu\partial_\mu\tilde{\phi} + (e^2\,A^\mu A_\mu - m^2)\,\tilde{\phi}\Bigr)\,\phi^* \label{fu3} \,.
\end{eqnarray}

From the equality of mixed partial derivatives as shown in Eq. (\ref{cau}), it follows that 
\begin{equation}
\eta_\nu\partial^{\nu}A_\mu + 2\, (\partial_0\eta_0)\,A_\mu
- A_\nu\,\partial^{\nu}\eta_\mu = \partial_\mu\lambda \,, \label{tit}
\end{equation}
allowing to re-express Eqs. (\ref{fu1}) and (\ref{fu2}) as 
\begin{eqnarray}
\frac{\partial\sigma_\mu}{\partial\phi} &=& -\tilde{\phi}^*\,\partial_\mu\partial_{0}\eta_0 + \partial_\mu\tilde{\phi}^* 
- i\,e\,A_\mu\,\tilde{\phi}^* \,, \nonumber \\
\frac{\partial\sigma_\mu}{\partial\phi^*} &=& -\tilde{\phi}\,\partial_\mu\partial_{0}\eta_0 + \partial_\mu\tilde{\phi} + i\,e\,A_\mu\,\tilde{\phi} \,, \nonumber
\end{eqnarray}
which has the solution 
\begin{equation}
\sigma_\mu = - (\partial_\mu\partial_{0}\eta_0)\,\phi^*\,\phi + (\partial_\mu\tilde{\phi}^* 
- i\,e\,A_\mu\,\tilde{\phi}^*)\,\phi + (\partial_\mu\tilde{\phi} 
+ i\,e\,A_\mu\,\tilde{\phi})\,\phi^* + \tilde{\sigma}_{\mu}(x) \,, \label{sig}
\end{equation}
where $\tilde{\sigma}_{\mu}(x)$ does not depend on $\phi, \phi^*$.

Plugging $\sigma_\mu$ from Eq. (\ref{sig}) into Eq. (\ref{fu3}) and considering the terms linear in $\phi^*$ and $\phi$, the  results are
\begin{eqnarray}
\Bigl(\partial^\mu\partial_\mu + 2\,i\,e\,A^\mu\partial_\mu + i\,e\,\partial^\mu A_\mu - (e^2\,A^\mu A_\mu - m^2)\Bigr)\,\tilde{\phi} &=& 0 \,, \nonumber\\
\Bigl(\partial^\mu\partial_\mu - 2\,i\,e\,A^\mu\partial_\mu - i\,e\,\partial^\mu A_\mu - (e^2\,A^\mu A_\mu - m^2)\Bigr)\,\tilde{\phi}^* &=& 0 \,, \nonumber
\end{eqnarray}
which are the Klein-Gordon equations obeyed by $\tilde{\phi}, \tilde{\phi}^*$. Therefore, the addition of particular solutions is an allowable Noether symmetry, reflecting the linearity of the problem. On the same trend, the independent term (not containing $\phi$ or $\phi^*$) gives $\partial^\mu\tilde\sigma_\mu = 0$, so that $\tilde\sigma_\mu = 0$ without loss of generality. 

In the continuation, the term proportional to $\phi^*\phi$ in Eq. (\ref{fu3}), taking into account Eq. (\ref{sig}), gives 
\begin{equation}
\Bigl(\partial^\mu\partial_\mu + 2\,(e^2\,A^\mu A_\mu - m^2)\Bigr)\,\partial_{0}\eta_0 + 2\,e^2\eta_\nu\,(\partial^\nu A_\mu)\,A^\mu - 2\,e^2\,A^\mu\partial_\mu\lambda = 0 \,.
\nonumber
\end{equation}
Using Eq. (\ref{tit}), the last equation can be reshaped as 
\begin{equation}
\Bigl(\partial^\mu\partial_\mu - 2\,(e^2\,A^\mu A_\mu + m^2)\Bigr)\,\partial_{0}\eta_0 + 2\,e^2\,A^\mu A_\nu\,\partial^\nu\eta_\mu = 0 \,,
\nonumber
\end{equation}
where the gauge function $\lambda$ was eliminated. From the identities shown in Eqs. (\ref{x1}) and (\ref{xx1}), it is possible to prove that $A^\mu A_\mu\,\partial_0\eta_0 = A^\mu A_\nu\,\partial^\nu\eta_\mu = 0$, so that 
\begin{equation}
\Bigl(\partial^\mu\partial_\mu - 2\,m^2\Bigr)\,\partial_{0}\eta_0 = 0 \,.
\nonumber
\end{equation}
We have already meet the same equation, see Eq. (\ref{eq4}). Following the same procedure as before, we obtain $\partial_{0}\eta_0 = 0$, which in turn imply $\partial_\mu\eta_\nu + \partial_\nu\eta_\mu = 0$, showing that the symmetry transformations constitute the Poincar\'e group. In comparison, the symmetry treatment of non-relativistic charged particle motion under external electromagnetic fields, including magnetic monopoles, is compatible with much more general transformations of the time variable \cite{Haas1, Haas2, Haas3}. In the relativistic case, the space and time variables entanglement allows only linear coordinate transformations as determined by the Poincar\'e generators. 

The external fields are not arbitrary. According to Eq. (\ref{tit}), they satisfy 
\begin{equation}
\eta_\nu\partial^{\nu}A_\mu - A_\nu\,\partial^{\nu}\eta_\mu = \partial_\mu\lambda \,, \label{tiit} \nonumber
\end{equation}
for some function $\lambda$. 
In other words, 
\begin{equation}
\eta_\nu (\partial^\nu A_\mu - \partial_\mu A^\nu) = \partial_\mu (\lambda - \eta^\nu A_\nu) + (\partial_\mu \eta_\nu + \partial_\nu \eta_\mu)\,A^\nu \,, 
\nonumber
\end{equation}
which is the same as 
\begin{equation}
\label{grad}
\eta^\nu F_{\nu\mu} = \partial_\mu\tilde{\lambda} \,,
\end{equation}
where 
\begin{equation}
\tilde{\lambda} = \lambda - \eta^\mu A_\mu \nonumber
\end{equation}
is a redefined arbitrary function. Equation (\ref{grad}) is in a manifestly covariant form. The last requirement is the verification of the homogeneous Maxwell equations, 
\begin{equation}
\label{homo}
\partial_\mu F_{\nu\alpha} + \partial_\nu F_{\alpha\mu} + \partial_\alpha F_{\mu\nu} = 0 \,. \nonumber
\end{equation}

\subsection{Examples}

It is worthwhile to consider illustrations of the compact equation (\ref{grad}).  

\subsubsection{Time-translations}

Supposing time-translations, one has $\lambda_0 = 1, \lambda_i = 0$. In this case, it is a simple matter to verify that Eq. (\ref{grad}) implies an electric field ${\bf E} = \nabla\tilde{\lambda}$, where $\tilde{\lambda} = \tilde{\lambda}({\bf r})$ is time-independent. Finally, Faraday's law shows that the magnetic-field is also time-independent. 

\subsubsection{Space-translations}

For translations along a particular axis (say, the $z-$direction), one has $\eta_0 = \eta_1 = \eta_2 = 0, \eta_3 = 1$. In this case, applying the symmetry condition (\ref{grad}), one has the electromagnetic field components
\begin{equation}
B_1 = F_{32} =  \partial^{2}\tilde{\lambda} \,, \quad B_2 = F_{13} = - \partial^{1}\tilde{\lambda} \,, \quad E_3 = F_{30} = - \partial_{0}\tilde{\lambda} \,, \nonumber
\end{equation}
where $\partial^{3}\tilde{\lambda} = 0$. Moreover, from the homogeneous Maxwell's equations it is found that $\partial^{3}E_1 = \partial^{3}E_2 = \partial^{3}B_3 = 0$. 

\subsubsection{Circularly Polarized Electromagnetic Field}

In an inverse approach, one might consider first the electromagnetic configuration, and then ask about Noether symmetries plugging the field into Eq. (\ref{grad}). For instance, for the right circularly polarized wave one has 
\begin{equation}
\label{a3}
{\bf A} = \frac{C_0}{\sqrt{2}} ({\boldsymbol \epsilon}\, e^{i\theta} + {\boldsymbol \epsilon}^* e^{-i\theta}) \,, \quad A_0 = 0 \,, \nonumber
\end{equation}
where $\omega, k$ are constants and $C_0$ is a slowly varying function of the phase
\begin{equation}
\label{phase}
\theta = k z - \omega t  \,, \nonumber
\end{equation}
while ${\boldsymbol \epsilon} = (\hat{x} - i \hat{y})/\sqrt{2}$ denotes the polarization vector, with the unit vectors $\hat{x}, \hat{y}$ perpendicular to the direction of light propagation. It is a simple matter to verify that the circularly polarized field admits Noether symmetries, satisfying Eq. (\ref{grad}) with $\eta_0 = k, \eta_1 = \eta_2 = 0, \eta_3 = \omega$, together with $\tilde{\lambda} = 0$. In this case, it should be noticed that this is not the unique solution to the system provided by Eq. (\ref{grad}). 

\subsubsection{Homogenous Static Magnetic Field}

As a final example, let us suppose an homogeneous static magnetic field
\begin{equation}
{\bf B} = B_0 \hat{z} \,, \nonumber
\end{equation}
where $B_0$ is a non-zero constant. What are the allowable electric fields, so that Noether symmetry exist? 

From Faraday’s law, one has $\nabla\times{\bf E} = 0$, so that ${\bf E} = - \nabla\,A_0$, where $A_0$ is the scalar potential. Therefore, 
$F_{i0} = - \partial^{i}A_0$. Moreover, for our magnetic field we have $F_{12} = - B_0, F_{23} = F_{31} = 0$. The Noether symmetry 
condition (\ref{grad}) reduces to four equations. For simplicity, limiting ourselves to the case $\eta_0 = 0$, these equations are given by
\begin{eqnarray}
\eta_{i}\partial^{i}A_0 &=& \partial\tilde{\lambda}/\partial t \,, \label{fffu1} \\
\eta_2 B_0 &=& \partial\tilde{\lambda}/\partial x \,, \label{fffu2} \\
- \eta_1 B_0 &=& \partial\tilde{\lambda}/\partial y \,, \label{fffu3} \\
0 &=& \partial\tilde{\lambda}/\partial z \,. \label{fffu4}
\end{eqnarray}
Since $\eta_0 = 0$ implies $a^0 = 0, R^{0i} = 0$, the general solution for Eqs. (\ref{fffu2}) and (\ref{fffu3}) is 
\begin{equation}
\label{fffu5}
\tilde{\lambda} = B_0 (a^1 y - a^2 x) + \frac{1}{2}\,B_0 R^{12} (x^2 + y^2) - B_{0} z (R^{23} x + R^{31} y) + g(z,t) \,,
\end{equation}
for some function $g(z,t)$.
In view of Eq. (\ref{fffu4}), from Eq. (\ref{fffu5}) we conclude that $R^{23} = R^{31} = 0$, and that $g = g(t)$. Finally, Eq. (\ref{fffu1}) gives 
\begin{equation}
\label{fffu6}
(a^1 + R^{12} y)\,\frac{\partial A_0}{\partial x} + (a^2 - R^{12} x)\,\frac{\partial A_0}{\partial y} + a^3\,\frac{\partial A_0}{\partial z} = - \dot{g}(t) \,.
\end{equation}
Just for simplicity, we will assume $a^3 = 0$. The results with $a^3 \neq 0$ are easily reachable, although of a less readable form. 

There are two classes of solutions for Eq. (\ref{fffu6}) with $a^3 = 0$. The first, for $R^{12} \neq 0$, can be more simply written taking $R^{12} = 1$ and $a^{1} = a^2 = 0$ without loss of generality, after a rescaling of $g$ and appropriate space translations, if necessary. In this case one has 
\begin{equation}
A_0 = \dot{g} \,\,{\rm arctan}\left(\frac{y}{x}\right) + \tilde{A}_{0}\left(\sqrt{x^2 + y^2}, z, t\right) \,, \nonumber
\end{equation}
where $\tilde{A}_0$ is an arbitrary function of the indicated arguments. Notice that the electric field found from the scalar potential is a single valued function. 

The second class of solutions for Eq. (\ref{fffu6}) is found for $R^{12} = 0$ and is given by 
\begin{equation}
A_0 = - \frac{1}{2}\,\dot{g}\left(\frac{x}{a^1} + \frac{y}{a^2}\right) + \tilde{A}_{0}(a^1 y - a^2 x, z, t) \,, \nonumber
\end{equation}
where $\tilde{A}_0$ is an arbitrary function of the indicated arguments and where $a^1 a^2 \neq 0$ was assumed, for simplicity.

\subsection{Conserved Currents for the Charged Scalar Particle under External Electromagnetic Fields}

Our conserved current reads
\begin{eqnarray}
J^\mu &=& \Bigl[\Bigl(D^\mu\phi\Bigr)\,\partial_\nu\phi^* + \Bigl(D^\mu\phi\Bigr)^{*}\,\partial_\nu\phi - \delta^{\mu}_{\,\nu}\,\cal{L}\Bigr]\,\eta^\nu \nonumber \\
&+& i\,e\,\lambda\,\Bigl[\phi\,\Bigl(D^\mu\phi\Bigr)^* - \phi^{*}\,\Bigl(D^\mu\phi\Bigr)\Bigr] \nonumber \\
&+& \phi\,\Bigl(D^\mu\tilde{\phi}\Bigr)^* + \phi^{*}\,\Bigl(D^\mu\tilde{\phi}\Bigr) - \tilde{\phi}\,\Bigl(D^\mu\phi\Bigr)^*  - \tilde{\phi}^{*}\,\Bigl(D^\mu\phi\Bigr) \,. \nonumber 
\end{eqnarray}
There are three components in the current: the external symmetries contribution, associated to $\eta^\mu$; the internal gauge symmetries associated to $\lambda$; the internal symmetries due to linearity, due to the superposition law for particular solutions 
$\tilde{\phi}, \tilde{\phi}^*$.  

\section{Conclusion}
\label{conclusion}

Adopting a systematic procedure, it was seen that the incorporation of symmetries requires the addition of adequate terms to the Lagrangian density. Already in the case of the complex scalar field, there is a global gauge symmetry, which becomes local when coupled to the electromagnetic field. In this context, gauge symmetries can be viewed as rotations in the internal space parametrized by the real and imaginary parts of the field \cite{Ryder}. For the vacuum electromagnetic field, we have a local gauge symmetry, which is in accordance with the causality principle. The accompanying linearity of the equations, is associated with a Noether conserved current not always emphasized in the literature. It was the main purpose of the present study, to pursue the calculations of Noether symmetries without {\it ab initio} assumptions. Such a procedure was shown to be useful for the clear identification of the internal symmetries due to linearity, in the case of the free theories (real and complex scalar field, vacuum Maxwell's equations). The conservation laws can be used to check the accuracy of numerical methods. It is expected, that the present systematic procedure could be more frequently applied, in both relativistic and non-relativistic studies. An example was applied, for the first time, to the case of a charged scalar particle under a general external electromagnetic field. The Noether symmetry condition, in this case, was reduced to the compact system provided by (\ref{grad}), which is manifestly gauge-invariant. ¨

\acknowledgments
Work partially supported by Con\-se\-lho Na\-cio\-nal de De\-sen\-vol\-vi\-men\-to Cien\-t\'{\i}\-fi\-co e Tec\-no\-l\'o\-gi\-co 
(CNPq).

\end{document}